\documentclass[preprint,12pt]{elsarticle}
\makeatletter
\def\ps@pprintTitle{} 
\makeatother

\usepackage{soul}

\usepackage{amssymb}
\usepackage{array}
\usepackage{rotating}
\usepackage{lineno}
\usepackage{algorithm}
\usepackage{algorithmic}
\usepackage{xcolor}
\usepackage{graphicx}
\usepackage{subcaption} 
\usepackage[margin=1in]{geometry}
\usepackage{amsmath}
\usepackage{amsthm}
\usepackage{booktabs}

\begin{document}

\begin{frontmatter}



\title{Effective Connectivity-Based Unsupervised Channel Selection Method for EEG} 


\author{Neda Abdollahpour\textsuperscript{a*}, N. Sertac Artan \textsuperscript{a}, Member, IEEE, Ian Daly\textsuperscript{b}, Mohammadreza Yazdchi\textsuperscript{c}, and Zahra Baharlouei\textsuperscript{d}}

\affiliation[1]{organization={Department of Electrical and Computer Engineering, New York Institute of Technology, New York, NY, USA}, 
            email={nabdol01@nyit.edu, nartan@nyit.edu}}

\affiliation[2]{organization={Electrical Engineering Department, University of Essex, Colchester, Essex, UK}, 
            email={i.daly@essex.ac.uk}}

\affiliation[3]{organization={Biomedical Engineering, University of Isfahan, Isfahan, Iran}, 
            email={yazdchi@eng.ui.ac.ir}}

\affiliation[4]{organization={Medical Image and Signal Processing Research Center, School of Advanced Technologies in Medicine, Isfahan University of Medical Sciences, Isfahan, Iran}, 
            email={zahra.bahar@res.mui.ac.ir}}

\begin{abstract}
    Analyzing neural data such as Electroencephalography (EEG) data often involves dealing with high-dimensional datasets, where not all channels provide equally meaningful information. Selecting the most relevant channels is crucial for improving computational efficiency and ensuring robust insights into neural dynamics. This study introduces the {\em Importance of Channels based on Effective Connectivity (ICEC)} criterion for quantifying effective connectivity (EC) in each channel. Effective connectivity refers to the causal influence one neural region exerts over another, providing insights into the directional flow of information. Using this criterion, we propose an unsupervised channel selection method that accounts for the intensity of interactions among channels. To evaluate the proposed channel selection method, we applied it to three well-known EEG datasets across four categories. The assessment involved calculating the ICEC criterion using five effective connectivity metrics: partial directed coherence (PDC), generalized PDC (GPDC), renormalized PDC (RPDC), directed transfer function (DTF), and direct DTF (dDTF). To focus on the effect of channel selection, we employed the Common Spatial Pattern (CSP) algorithm for feature extraction and a Support Vector Machine (SVM) for classification across all participants. Results were compared with other CSP-based methods. The evaluation included comparing participant-specific accuracies with and without the proposed method across five effective connectivity metrics. The results showed consistent performance improvements and a significant reduction in the number of selected electrodes for all participants.Compared to state-of-the-art methods, our approach achieved the highest accuracies: 82\% (13 out of 22 channels), 86.01\% (29 out of 59 channels), and 87.56\% (48 out 0f 118 channels) across three datasets. 
    \textit{(This paper has been accepted for publication in the \textit{Journal of Medical Signals \& Sensors} and will appear soon.)}




\end{abstract}




\end{frontmatter}



\section{Introduction}
\label{Intro}

Electroencephalograms (EEGs) play a crucial role in measuring brain neural activities across various applications, such as brain-computer interfaces (BCIs) and the mental disorders diagnosis \cite{1,2,3}. Compared to other neuroimaging techniques, EEG is cost-effective, user-friendly, and its applications in biomedical fields, including the diagnosis and treatment of brain diseases, are rapidly expanding \cite{4,5}. 

An effective channel selection technique is a critical step in EEG signal processing, as it not only reduces data dimensionality and computational complexity but also addresses practical challenges. For example, in motor imagery (MI) tasks designed for paralyzed individuals, efficient channel selection can help mitigate issues such as user fatigue during training \cite{6}. Numerous channel selection methods have been proposed to date, which can broadly be categorized into three groups: filter-based techniques \cite{7,8,9}, wrapper-based methods \cite{9,11}, and hybrid techniques that combine elements of both approaches \cite{11,12,13,14,15}.

Wrapper-based methods evaluate a subset of potentially useful EEG channels by assessing their impact on classification performance through iterative model training and validation. This approach ensures that only the most informative channels are retained, optimizing both accuracy and computational efficiency. While these methods can yield high accuracy, they are prone to overfitting and are computationally expensive \cite{11}. In contrast, filter-based techniques are faster, scalable, and do not require classifiers. However, their simplicity comes at the cost of lower accuracy compared to wrapper-based methods \cite{11}.

There are some EEG channel selection methods that were proposed during previous years to enhance the performance of EEG-based problems. For example, in 2020, a method called bispectrum-based EEG electrode selection (BCS-CSP) used the sum of logarithmic amplitudes (SLA) and the first-order spectral moment (FOSM) features extracted from bispectrum analysis for this purpose \cite{29}.  In 2018, a subject-specific multivariate empirical mode decomposition (MEMD) based filtering method (SS-MEMDBF) was proposed. In this method, the statistical measure, mean frequency, was used to automatically filter the MIMFs to obtain enhanced EEG signals that could represent MI-related brainwave modulations over \(\mu\) and \(\beta\) rhythms \cite{33}. 

Reza et al.~\cite{34} proposed a method called Improved Binary Gravitation Search Algorithm (IBGSA) in which the logarithmic power of each channel in the time domain and the wavelet domain features mean, mode, median, variance, and standard deviation were used to select EEG channels. Qiu et al.~\cite{30} reduced the computation time of the Sequential Floating Forward Selection (SFFS) method, which was slow when number of features is large. This improved SFFS method reduced the total computation time, by selecting or removing several channels in every iteration.  An earlier study proposed a novel sparse common spatial pattern (SCSP) algorithm for EEG channel selection. In that study, channel selection is formulated as an optimization problem to select the least number of channels constrained by the classification accuracy \cite{31}. It is worth mentioning that the focus of the aforementioned studies was on MI tasks.

In most cases, EEG sensor selection methods are tailored to specific tasks, taking into account the brain regions associated with those tasks. For instance, the motor cortex is highly active during motor imagery (MI) tasks, the visual cortex is involved in visual processing, and the prefrontal cortex (PFC) plays a key role in decision-making tasks \cite{10, abdollahpour2022eeg}. However, acquiring labeled trials for certain tasks, such as MI, can be both time-consuming \cite{16} and physically fatiguing for participants, particularly for individuals with movement disabilities, such as those with spinal cord or brain injuries. Consequently, researchers are increasingly exploring novel EEG sensor optimization techniques that are independent of specific tasks or the need for labeled trials.

In this paper, we propose an unsupervised method for selecting EEG electrodes based on effective connectivity. To achieve this, we introduce a criterion called the \textit{Importance of Channels based on Effective Connectivity} (ICEC), which quantifies the intensity of effective connectivity for each electrode. Using this criterion, we can identify and select an appropriate subset of electrodes. 

The novelty of this method lies in its simplicity and versatility. By focusing on enhancing classification accuracy while significantly reducing the number of EEG channels, it effectively lowers the dimensionality of the data, making it computationally efficient yet highly effective in improving performance. As an unsupervised approach, it does not require labeled data, potentially making it broadly applicable across various modalities and tasks. Additionally, it supports visualization and is adaptable to different states, such as resting-state and event-related tasks, further enhancing its utility in diverse neuroimaging applications. 




\section{MATHEMATICAL CONCEPTS}
\label{math}

A statistical concept of causality is provided by Granger causality. According to this concept, if a signal \(Y_1\) "Granger-causes" (or "G-causes") a signal \(Y_2\), then past values of \(Y_1\) should contain information that is helpful to predict \(Y_2\) above and beyond the information of the past values of \(Y_2\) alone. The mathematical formulation of the method is based on linear regression modeling of stochastic processes \cite{37}. 
The evaluation of causal interactions among various brain regions is fundamental to elucidating its functional operations. A promising way to identify such links is using multivariate autoregressive (MVAR) modeling methods, which are applied to multisite electrophysiological recordings. Indeed, they can estimate the extent to which past activity in one or more brain regions can predict current activity in another while taking into account the mediating effects of other regions \cite{17, 21}.

\subsection{Effective connectivity metrics}
An extension of Granger causality is PDC, which normalizes terms in the frequency domain by the total outflow at a node. DTF, on the other hand, normalizes frequency-domain terms by the total inflow at a node. This is an alternative causal model that normalizes directed effects by the sum of the transfer functions going into the node, as opposed to the sum of the transfer functions going out of the node as measured by the PDC \cite{17, 25}. Given the MVAR model in Eq. (1) with a model order of $\rho$, where $A(f)$ is the system matrix (Eq. (2)), and $E$ shows the residuals. Eq. (3) shows the Fourier Transform of the data, and $H(f)$ is the transfer function.

\begin{align}
X(t) &= \sum_{k=1}^{\rho} A^{(k)}(t) X(t-k) + E(t) \tag{1} \\
A(f, t) &= -\sum_{k=0}^{\rho} A^{(k)}(t) e^{-i 2 \pi f k}; \quad A^{(0)} = I \tag{2} \\
X(f, t) &= A(f, t)^{-1} E(f, t) = H(f, t) E(f, t) \tag{3}
\end{align}

Then, the DTF can be formulated as in Eq. (4), where $M$ is the number of variables \cite{17, 26, 27, 28}.

\begin{equation}
\eta_{ij}^2(f) = \frac{|H_{ij}(f)|^2}{\sum_{f} \sum_{k=1}^M |H_{ik}(f)|^2}; \quad M > 2 \tag{4}
\end{equation}

From DTF, dDTF can be calculated as in Eq. (5), where \(P_{ij}(f)\) is Partial Coherence and \(\eta_{ij}^2(f)\) is  Full-Frequency DTF (ffDTF). PDC ($\Pi$) and GPDC ($\pi$) are determined by Eq. (6) and (7), respectively.
\begin{equation}
\delta_{ij}^2(f) = \eta_{ij}^2(f) P_{ij}^2(f) \tag{5}
\end{equation}
\begin{equation}
\Pi_{ij}^2(f) = \frac{|A_{ij}(f)|^2}{\sum_{k=1}^M |A_{kj}(f)|^2} \tag{6}
\end{equation}

\begin{align*}
\overline{\pi}_{ij}(f) &= \frac{\frac{1}{\sum_{ii}} A_{ij}(f)}{\sqrt{\sum_{k=1}^{M} \frac{1}{\sum_{ii}^2} |A_{kj}(f)|^2}}\tag{7} \\
0 &\leq |\overline{\pi}_{ij}(f)|^2 \leq 1 \; ; \; \sum_{j=1}^{M} |\overline{\pi}_{ij}(f)|^2 = 1
\end{align*}
where \(\Sigma\) is the noise covariance matrix. RPDC ($\lambda$)is calculated by Eq. (8).
\begin{equation}
\lambda_{ij}(f) = Q_{ij}(f) \times V_{ij}(f)^{-1} Q_{ij}(f) \tag{8}
\end{equation}
\begin{equation}
Q_{ij}(f) = 
\begin{pmatrix}
\text{Re}[A_{ij}(f)] \\
\text{Im}[A_{ij}(f)]
\end{pmatrix} \tag{9}
\end{equation}
\begin{equation}
V_{ij}(f) = \sum_{k,l=1}^{\rho} R_{jj}^{-1}(k, l) \Sigma_{ii} Z(2\pi f, k, l) \tag{10}
\end{equation}
\begin{equation}
Z(\omega, k, l) = 
\begin{pmatrix}
\cos(\omega k) \cos(\omega l) & \cos(\omega k) \sin(\omega l) \\
\sin(\omega k) \cos(\omega l) & \sin(\omega k) \sin(\omega l)
\end{pmatrix} \tag{11}
\end{equation}
where \(R\) is the \([ (M\rho)^2 \times (M\rho)^2 ]\) covariance matrix of the VAR\([\rho]\) process, a vector autoregression of order \(\rho\) \cite{17, 44}.

\subsection{Model order}
Model order (\(\rho\)) assumes a pivotal role in fitting an MVAR model. Indeed, the quality of model fitting is contingent on the order of the model. Model orders that are too small or too large can result in spectra that lack the necessary detail or create spurious maxima in the corresponding spectrum, respectively \cite{17, 28}.  To determine the optimum order of the model different criteria such as Akaike’s information criterion (AIC), Hannan–Quinn’s criterion, and Bayesian–Schwartz’s information criterion (BIC) have been proposed\cite{17}. In this study, we used the AIC method, which is the most preferred method \cite{17}, to find the proper model order. Akaike’s information criterion is calculated as
\begin{equation}
\text{AIC}(\rho) = \ln \big(|\det (V)|\big) + \frac{2}{N} \rho K^2 \tag{12}
\end{equation}

where \(V\) is the noise and \(E(t)\) the covariance matrix. Typically, the first term of the criterion depends on the estimated residual variance \(V(\rho)\) for a given \(\rho\), and the second term is a function of model order \(\rho\), the number of channels \(K\), and the number of data points \(N\) \cite{17, 26}.

\subsection{VAR model validation}
There are a number of criteria we can use to determine whether we are fitting our VAR model appropriately. Three commonly used categories of tests are Whiteness tests for checking the residuals of the model for serial and cross-correlation, Consistency tests to evaluate whether the model generates data with the same correlation structure as the real data, and the Stability test to check the stability/stationarity of the model \cite{17}. In this study, we selected a model order for the Multivariate VAR model that satisfied the stability criterion for all samples, ensuring it was specific to each participant. This means that for each participant, we determined the appropriate model order that maintained the stability of the MVAR model across all their respective data samples. By doing so, we ensured that the fitted MVAR models were both participant-specific and stable, which is crucial for accurate analysis of brain connectivity.

\section{METHODOLOGY}
\label{method}
In this section, we introduce our proposed EEG channel selection method. We describe datasets, the main steps of the proposed method and its formulation. 

\subsection{Data Descriptions}
\begin{figure}[htbp]
\centerline{\includegraphics[width=0.55\columnwidth]{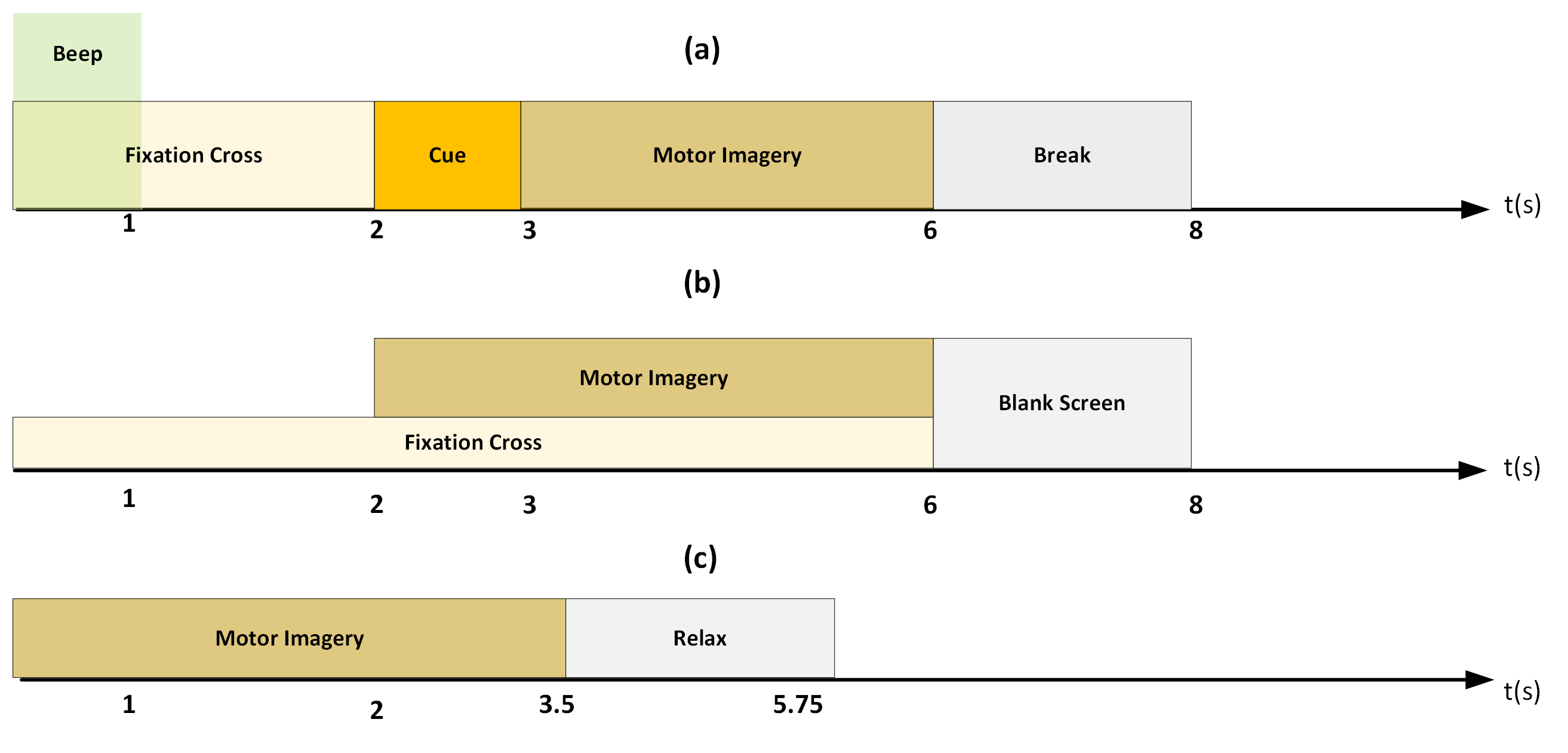}}
\caption{\textbf{Timeline of each trial for dataset1 (a), dataset2 (b) and dataset3 (c).} }
\label{data}
\end{figure}

\textbf{Dataset 1:} We use dataset 2a from the BCI-Competition IV \cite{22}, 2008. This is a four-class MI BCI dataset. It contains EEG signals from 9 healthy participants who performed 4 different MI tasks (imagining the movement of the left hand, right hand, leg, and tongue). Each dataset contains 288 labeled trials (training data), 72 trials per class, and 288 test trials. A detailed timeline of a single MI trial within this dataset is shown in Figure~\ref{data}(a). In this dataset, 22 EEG electrodes were used to record brain signals, and the sample rate of the EEG is 250 Hz. The position of the electrodes is based on the international EEG 10-20 systems.
\textbf{Dataset 2:} We use dataset 1 from the BCI-Competition IV \cite{23}. This is a two-class MI BCI dataset. It contains MI EEG signals from 7 participants. However, we eliminated participants “c”, “d”, and “e” as they were generated artificially and assessed only the remaining participants. During the first two runs, arrows pointing left, right, or down were presented as visual cues on a computer screen for 4 s. During this time span, each participant was instructed to perform the cued MI task, followed by 4 runs 
with unequal time periods of MI trials ranging from 1.5 to 8 s. The timeline of every MI trial is shown in Fig. \ref{data}(b). In this dataset, 59 EEG electrodes were used to record brain signals, and the sample rate of the signals is 1000 Hz, which was downsampled to 100 Hz.

\textbf{Dataset 3:} We use dataset IVa, BCI-Competition III \cite{24}. This is a two-class MI BCI dataset with small training sets. It contains MI EEG signals recorded from five healthy participants $aa$, $al$, $av$, $aw$ and $ay$. Each participant performed 280 MI tasks containing train and test trials. The number of train and test trials is not equal for all participants. In each run, there are 3.5-second visual cues for the three MI tasks the participant is asked to perform: (L) left hand, (R) right hand, and (F) right foot. In this dataset, 118 EEG electrodes were used to record brain signals, and the sample rate of the signals is 100 Hz. The timeline of every MI trial is shown in Fig. \ref{data}(c).

\subsection{Proposed Method}
\subsubsection{Acquiring 4-D Effective Connectivity matrices}
To acquire 4-dimensional effective connectivity matrices, we used EEGLAB Toolbox \cite{39} (version 2021.1) and MATLAB (version R2018b). The only preprocessing applied to the datasets was band-pass filtering the signal to the 1-40 Hz interval using a fifth-order Butterworth filter. In this study, we used 1000, 750, 400, and 250 time points for dataset 1 (4-class), dataset 1 (2-class), dataset 2, and dataset 3, respectively. These were selected to align with the specific requirements and structures of each dataset, ensuring that the selected time range accurately captured the MI tasks within each experimental framework. Then, the model was preprocessed by normalizing trials across time (method: ensemble) for each participant. 

In this study, the model order was determined using the AIC criterion. The selected model order was adjusted to ensure that the model stability test was satisfied for all participants \cite{17}. Using this chosen model order, the model was fitted with a window length of 0.5 seconds and a window step size of 0.03 seconds for all participants. Finally, connectivity measures were computed in the last stage across the frequency range of 1--40 Hz.

In the next stage, we calculated ICEC measures for 7 different frequency ranges. In this study, we acquire 5 effective connectivity matrices (DTF, dDTF, PDC, GPDC, and RPDC) and use them as 4-dimensional input maps for our proposed method.

\subsubsection{Calculating ICEC}
\begin{figure}[htbp]
\centerline{\includegraphics[width=0.9\columnwidth]{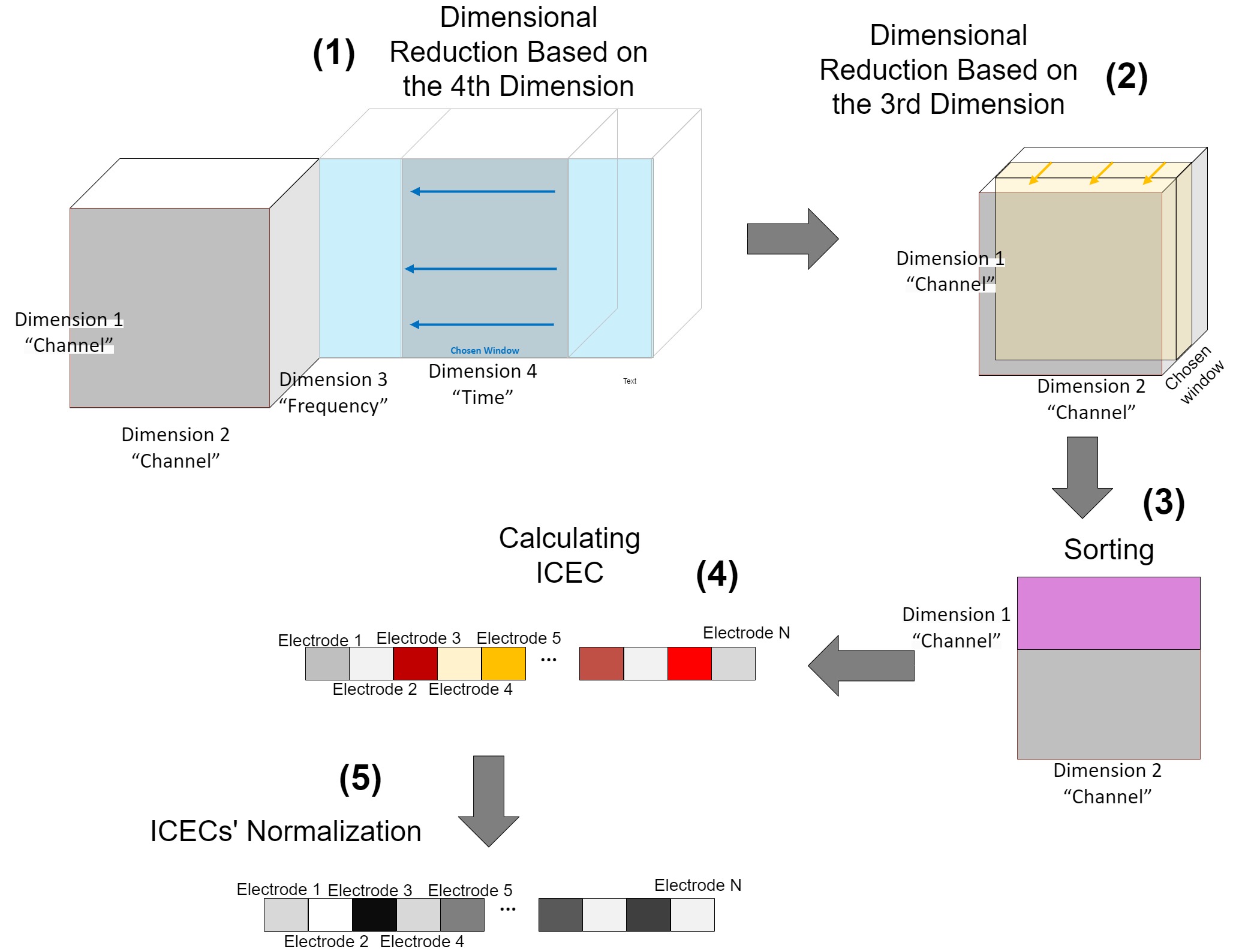}}
\caption{Five major steps for calculating the importance of each electrode: the first two steps reduce the dimensions of a 4D effective connectivity matrix for a participant, followed by sorting, ICEC calculation, and normalization. }
\label{main}
\end{figure}

Figure~\ref{main} outlines the 5 main steps for calculating the ICEC amount for each channel. As can be seen from the illustrations, in the first two steps, the dimensions of the input are reduced. In these stages, we use specific windows that determine the borders of the third (frequency) and the fourth (time) dimensions, respectively. 

In this study, we use seven different windows for the third dimension (7 different frequency ranges). We use all time samples without applying a specific window. The result of the aforementioned stages is a square matrix (\(N_{\text{ch}} \times N_{\text{ch}}\)) where \(N_{\text{ch}}\) is the total number of electrodes for each task. Every value within this matrix can also be calculated by Eq. (13) using Algorithm 1.
\begin{equation}
C_{i \times j} = \frac{\sum\limits_{f=f_{\text{min}}}^{f_{\text{max}}} \sum\limits_{w=w_{\text{min}}}^{w_{\text{max}}} \text{con}_{i \times j \times f \times w}}{(w_{\text{max}} - w_{\text{min}})(f_{\text{max}} - f_{\text{min}})} \tag{13}
\end{equation}
As for the \(C_{i \times j}\) amounts, when \(\text{con}\) is a 4-dimensional effective connectivity matrix, \(C_{i \times j}\) is a measure that depicts the mean value of interactions between node \(i\) and \(j\). Calculating \(C_{i \times j}\) for all channels gives us a square matrix.
After sorting \( C_{i \times j} \) amounts within the matrix and summing up the 30 percent of highest ones for each electrode, we possess ICEC for each EEG channel as a factor to evaluate the importance of each electrode. The 30 percent was chosen by trial and error in this study. After normalizing the figures, in the last stage, we selected some subsets of electrodes considering their highest amounts of ICECs in a circle with the usage of a classifier to evaluate their suitability for aiding EEG decoding.

\begin{algorithm}
\centering
\label{algorithm}
\caption{\textit{Effective Connectivity-Based Channel Selection}}
\begin{algorithmic}[1]
\STATE Read 4-D Effective Connectivity (labeled ``con'') \\
\hspace{2em} $con \in \mathbb{R}^{N_{Ch} \times N_{Ch} \times N_f \times N_w}$ \\
\hspace{2em} $N_f =$ Number of specific frequencies chosen \\
\hspace{2em} $N_w =$ Number of windows (time points) \\
\hspace{2em} $N_{Ch} =$ Number of channels
\FOR{$i = 1 : N_{Ch}$} 
    \FOR{$j = 1 : N_{Ch}$} 
        \WHILE{$i \neq j$}
            \STATE Calculate $C_{i,j}$ according to Eq. (13)
            \STATE Construct matrix ``$Z_s$'' containing $C_{i,j}$ measures for all nodes
            
        \ENDWHILE
    \ENDFOR
\ENDFOR
\STATE Sort $Z_s$ measures based on the first dimension \\
\hspace{2em} $Z_s \in \mathbb{R}^{N_{Ch} \times N_{Ch}}$
\STATE Calculate the sum of the first 30 percent amounts of $Z_s$ (labeled ``M'')
\STATE Sort $M$ from the maximum amount to the minimum
\STATE \textit{ICEC:} Calculate the index and amount of \textit{ICEC} for each electrode \\
\hspace{2em} \textbf{Index:} Selected electrodes \\
\hspace{2em} \textbf{Amount:} Weight of each node (\textit{Normalized ICEC})
\end{algorithmic}
\end{algorithm}

\subsubsection{Frequency ranges}
In the process of calculating \( C_{i \times j} \), we consider 7 different frequency ranges to assess the proposed model for each user. They are Theta (4--7 Hz), Mu (8--12 Hz), Low-beta (13--15 Hz), High-beta (18--30 Hz), 29--40 Hz, 8--30 Hz, and 1--40 Hz. There is also the possibility of selecting a particular period of time within the timeframe of a trial; however, in this study, we use all windows (time points) without any selection.

\subsubsection{Time points and AMVAR details}
In this study, we used adaptive multivariate autoregressive (AMVAR) models trained with the Vieira-Morf algorithm \cite{17}, for all participants. The model order amounts range from 15 to 40, depending on the size of the dataset selected for each user. In total, we use window lengths of 0.5 s and a window step size of 0.03 s for all users in the model fitting step. 

The number of windows and time points (the 4th dimension of the 4-D effective connectivity matrices) are 90, 67, and 117 for all participants from dataset 1, 2, and 3, respectively (2-class tasks). The time points for dataset 1, considering all classes (a 4-class task), are also 126 for all 9 participants.

\subsubsection{4-D effective connectivity matrices}

In this study, we calculate 5 types of effective connectivity for each user to assess the proposed method and also to make a comparison of their effectiveness. The 4-D effective connectivity matrix dimensions are “to,” “from,” “frequency range,” and “time,” respectively. As the “to” and “from” dimensions are EEG channels, the 3rd dimension indicates the ranges of frequencies in which the interactions among nodes are evaluated. The 4th dimension also shows each aforementioned window as a time point.

\subsubsection{Channel Selection Using ICEC}

As can be seen within the algorithm, after extracting a user-dependent 4-D matrix, ICEC measures are calculated for each EEG channel in 7 different frequency ranges. After sorting these criteria measures and normalizing them, we select a subset of EEG channels with the highest amounts of ICEC. We evaluate our model by selecting electrodes ranging from 5 to the maximum of 22, 59, and 118 electrodes for datasets 1, 2, and 3, respectively. 

\begin{figure}[htbp]
\centerline{\includegraphics[width=0.9\columnwidth]{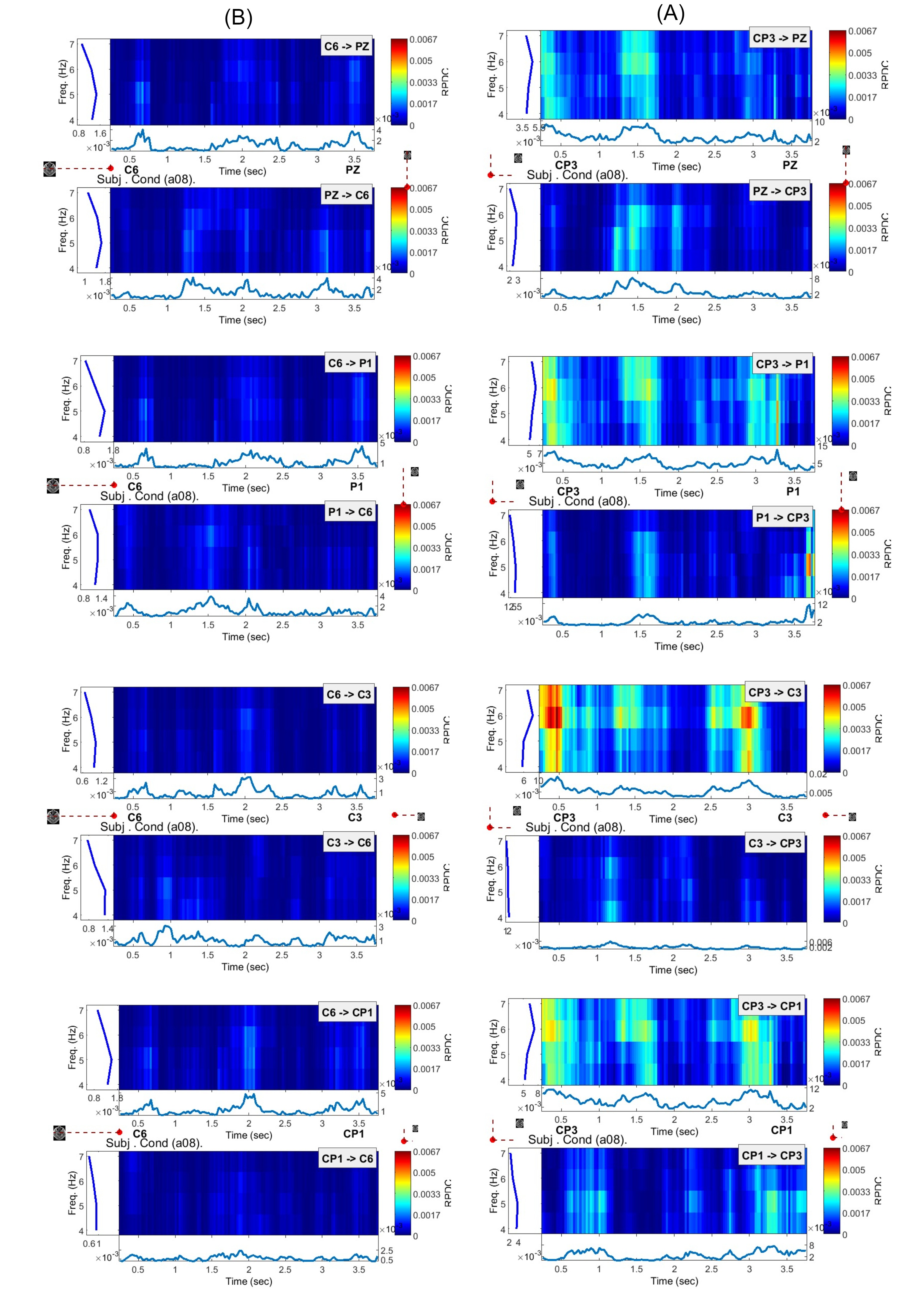}}
\caption{Comparison of RPDC interactions for participant ``a08'' in Dataset 1 between  electrodes (A) C6 (low ICEC) and (B) CP3 (high ICEC) with four other electrodes (P1, CP1, PZ, and C3)}
\label{low and high}   
\end{figure}

Figure~\ref{low and high} depicts interactions based on RPDC measures, between two specific EEG electrodes (CP3 with a high ICEC amount and C6 with a low ICEC amount) and 4 other electrodes (P1, CP1, PZ, and C3) separately, (a) and (b). We chose these channels randomly to compare their interactions with both a channel with a high ICEC value and a channel with a low ICEC value.

\subsection{Whole Framework}

\begin{figure}[htbp]
\centerline{\includegraphics[width=0.9\columnwidth]{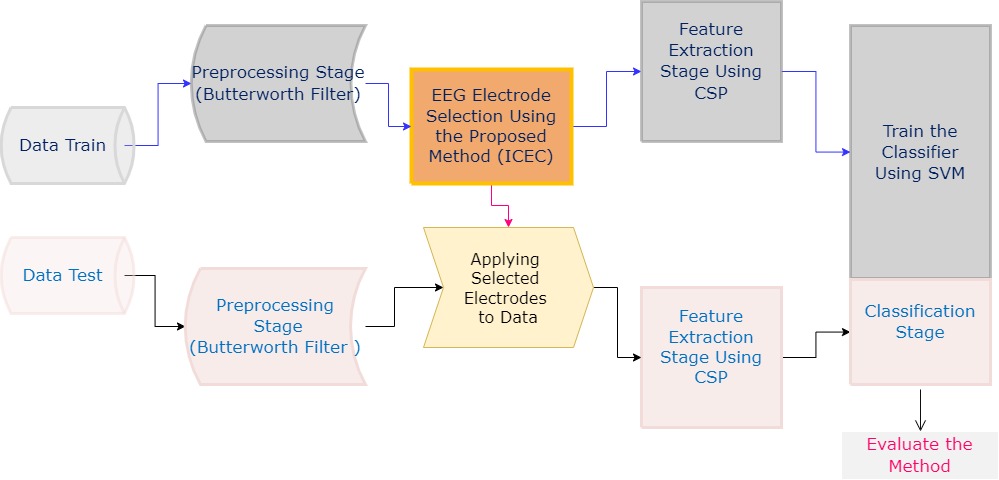}}
\caption{Main framework for evaluating the proposed EEG channel selection method using CSP in the feature extraction stage and SVM for classifying.}
\label{main1}   
\end{figure}

The framework for evaluating the proposed EEG Channel Selection method for all datasets is the same and is shown in Figure~\ref{main1}. In the first stage, the preprocessing stage, the data is segmented into training and testing sets and filtered by a band-pass Butterworth filter with order 3 and between 8--30 Hz. Dataset 1 was segmented between 625--1374 time points for all users and for both 2-class and 4-class categories. In addition, 2.5 seconds (containing 250 samples) are segmented for all users of dataset 2, while 4 seconds (400 samples) are used for dataset 3.

In the feature extraction stage, we use binary Common Spatial Patterns (CSP) for 2-class tasks and multi-class CSP for dataset 1. The feature used in this study is the variance of data distribution after being affected by special weights, acquired by CSP. In the classification stage, we apply a binary Support Vector Machine (SVM) to classify all 2-class problems and multi-class SVM for dataset 1. We use the RBF kernel in the SVM classifier and the One-versus-the-Rest technique in both multi-class CSP for the feature extraction stage and multi-class SVM for classifying multi-class datasets. 

It is important to highlight that for each frequency range, all stages of evaluation were conducted using the corresponding specific frequency range. For instance, in the case of the mu frequency range, the ICEC method was applied to the filtered data in this range (mu). Subsequently, feature extraction and classification were also performed on the filtered datasets.

\section{RESULTS AND DISCUSSIONS}
\label{results}
\subsection{Impact of Channel Selection on Classification Performance Across Subjects }

In this study, we use 5 effective connectivity measures (DTF, dDTF, PDC, GPDC, and RPDC) to calculate the ICEC criterion. This criterion is also assessed through 7 different frequency ranges: theta, mu, low-beta, high-beta, gamma, 8--30 Hz, and 1--40 Hz for each participant. 

\begin{figure}[htbp]
\centerline{\includegraphics[width=0.65\columnwidth]{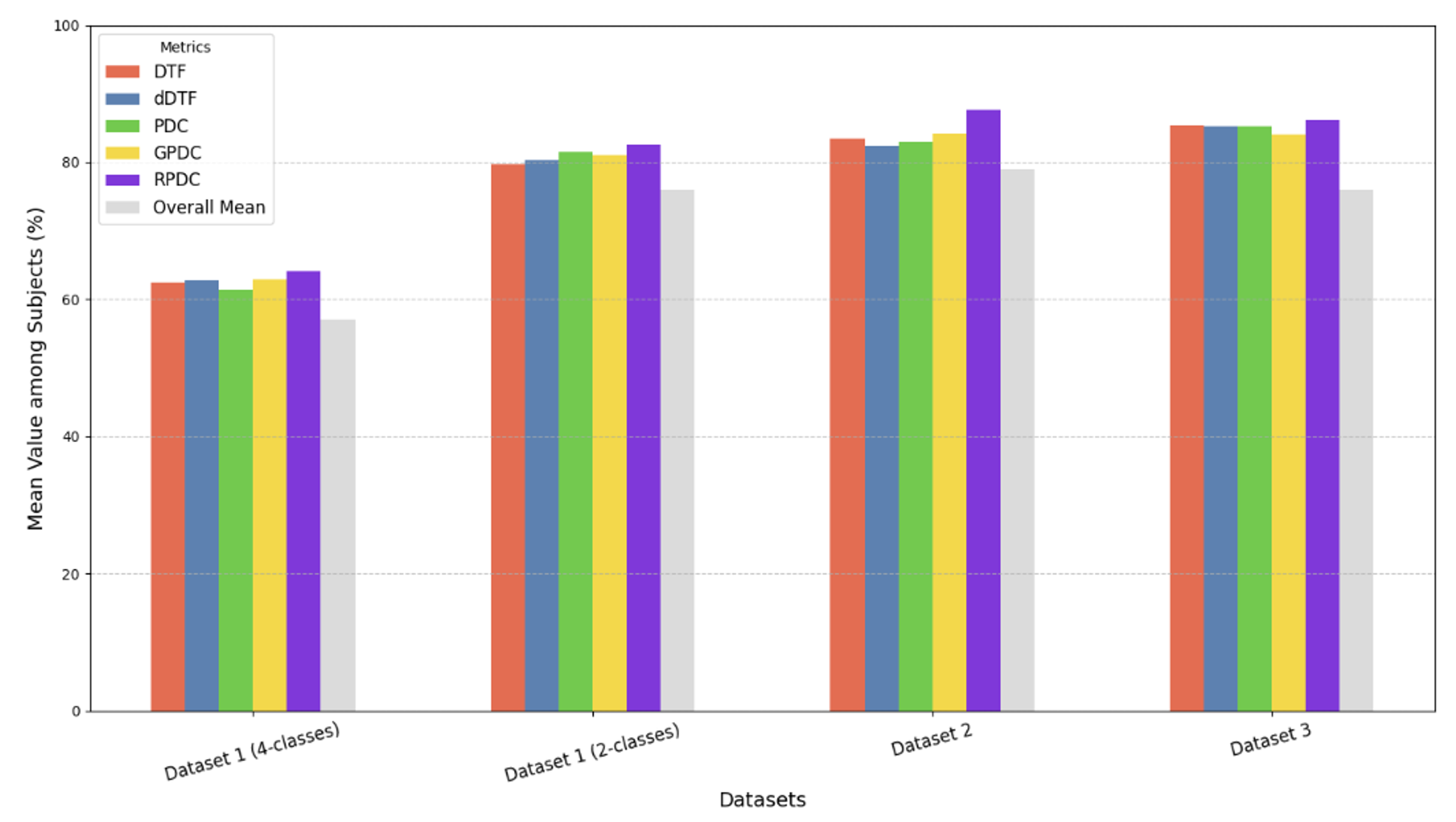}}
\caption{Comparison of mean accuracy values across subjects for different datasets (Dataset 1: 4-class and 2-class tasks, Dataset 2, and Dataset 3). The bars represent the performance metrics using various effective connectivity measures (DTF, dDTF, PDC, GPDC, RPDC) for channel selection and the overall mean accuracy. }
\label{bar chart}   
\end{figure}

Figure~\ref{bar chart} presents the average accuracies across subjects for the four datasets evaluated in this study using the ICEC criterion. As shown in the figure, channel selection based on all types of effective connectivity (EC) metrics led to improved performance across datasets, with RPDC demonstrating particularly strong results. For instance, for dataset 2, the accuracy values for RPDC consistently exceeded 86\%, while other metrics, such as GPDC and PDC, achieved accuracies in the range of 80--85\%. These findings underscore the effectiveness of the proposed method in enhancing classification performance through EC-based channel selection, with RPDC making the most significant contribution.

\subsection{Illustrations of selected channels using normalized ICEC}

\begin{figure}[htbp]
\centerline{\includegraphics[width=0.9\columnwidth]{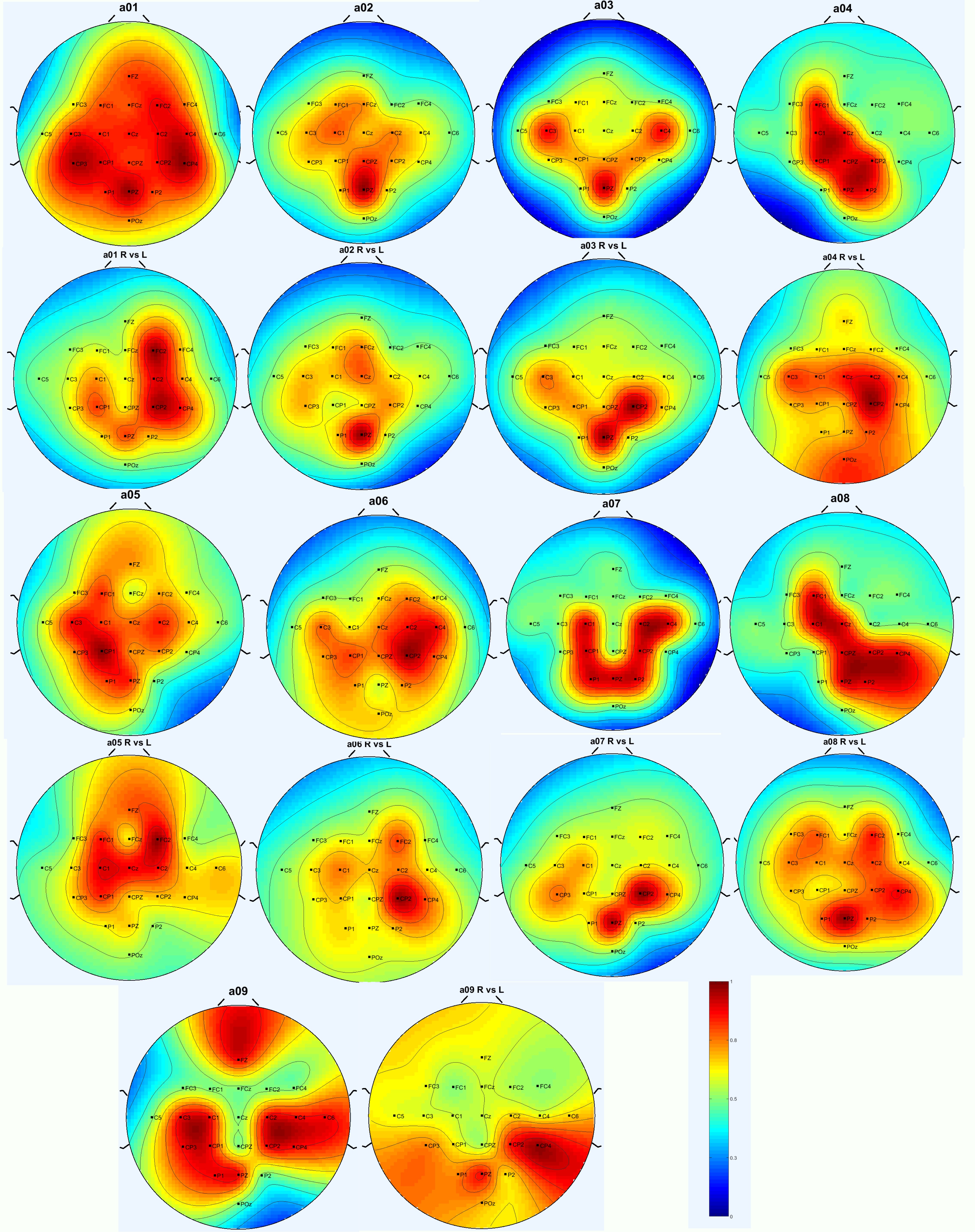}}
\caption{Two different topographies of selected channels using our proposed method for each participant (from Dataset 1) are shown. For each participant, the top illustration represents the importance of EEG channels in the 4-class task (imagination of the movement of the left hand, right hand, tongue, and foot), while the bottom illustration represents the importance of EEG channels in the 2-class motor imagery (MI) task (imagination of the movement of the left hand and right hand) with the exception of subject 9.
}
\label{plot1}   
\end{figure}

\begin{figure}[htbp]
\centerline{\includegraphics[width=0.9\columnwidth]{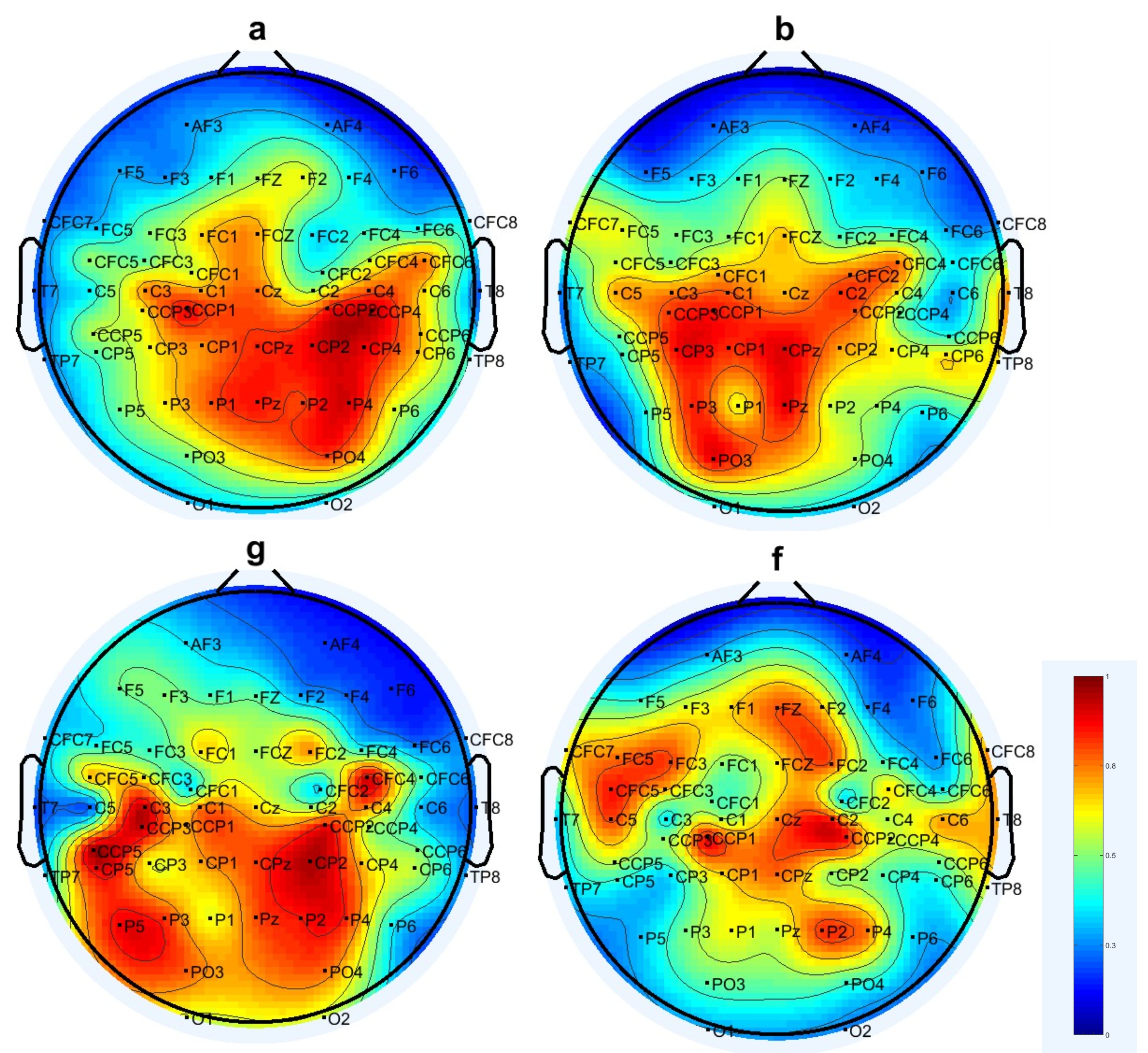}}
\caption{Topographical maps of channels selected from Dataset 2 with 59 EEG electrodes are shown. The color gradient, transitioning from red to blue, indicates decreasing prominence of the channels. To calculate ICEC for participants (a, b, f, and g), RPDC was applied in the frequency range of 29--40 Hz, GPDC in the low-beta range, dDTF in the mu range, and RPDC in the high-beta range, respectively.
}
\label{plot2}   
\end{figure}

\begin{figure}[htbp]
\centerline{\includegraphics[width=1\columnwidth]{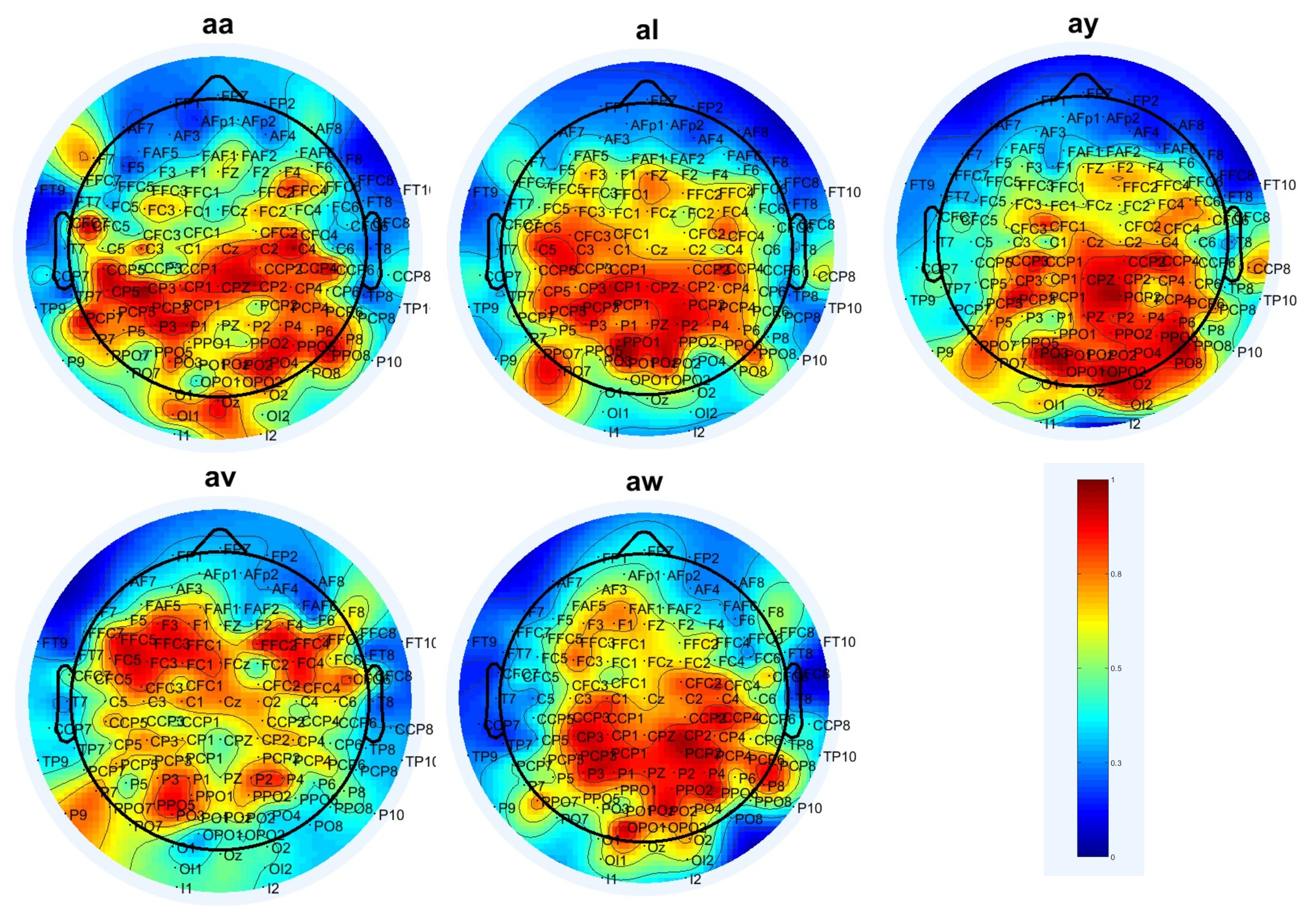}}
\caption{Topographical maps of channels selected from Dataset 3 with 118 EEG electrodes are shown. The color gradient, transitioning from red to blue, indicates decreasing prominence of the channels. To calculate ICEC for participants (aa, al, av, aw, and ay), we used dDTF in the high-beta frequency band, DTF in theta, RPDC in mu, DTF in 29--40 Hz, and RPDC in 29--40 Hz, respectively.
}
\label{plot3}  
\end{figure}

Figures~\ref{plot1}, \ref{plot2}, and \ref{plot3} illustrate the topographic maps of electrodes selected from datasets 1, 2, and 3, respectively. The Fig.\ref{plot3} presents topographical maps of EEG channels selected from Dataset 3, showcasing the spatial distribution of effective connectivity (ICEC) values for participants ``aa,'' ``al,'' ``av,'' ``aw,'' and ``ay.'' A color gradient is used to represent the prominence of the channels, transitioning from red (higher prominence) to blue (lower prominence). The maps highlight the regions of the scalp with the most significant effective connectivity, with each participant's analysis corresponding to a specific combination of effective connectivity metrics and frequency bands: dDTF in the high-beta band, DTF in the theta band, RPDC in the mu band, and RPDC or DTF in the 29--40 Hz band. These visualizations emphasize the individual variability in connectivity patterns and the distinct regions of neural activity identified by the proposed ICEC criterion, providing insights into the key channels that contribute most to classification accuracy. The Figures~\ref{plot1} and\ref{plot2}  also illustrate topographical maps of selected EEG channels from Datasets 1 and 2. 

Two distinct topographies of selected channels in Figure~\ref{plot1} using our proposed method are presented for each participant from Dataset 1. The top illustration for each participant depicts the importance of EEG channels in the 4-class MI task (imagining the movement of the left hand, right hand, tongue, and foot), while the bottom illustration represents the importance of EEG channels in the 2-class MI task (imagining the movement of the left hand and right hand), with the exception of subject 9. Comparing the channels with highest ICEC values shows that similar brain regions are active in both the 2-class and 4-class tasks. However, in the 4-class task, these regions are more extended, indicating a broader involvement of the brain when the number of imagined movements increases. This suggests that while certain regions remain consistently important across tasks, the complexity of the 4-class task recruits additional areas, reflecting the increased cognitive demand associated with distinguishing between more motor imagery classes.


\begin{figure}[htbp]
\centerline{\includegraphics[width=1\columnwidth]{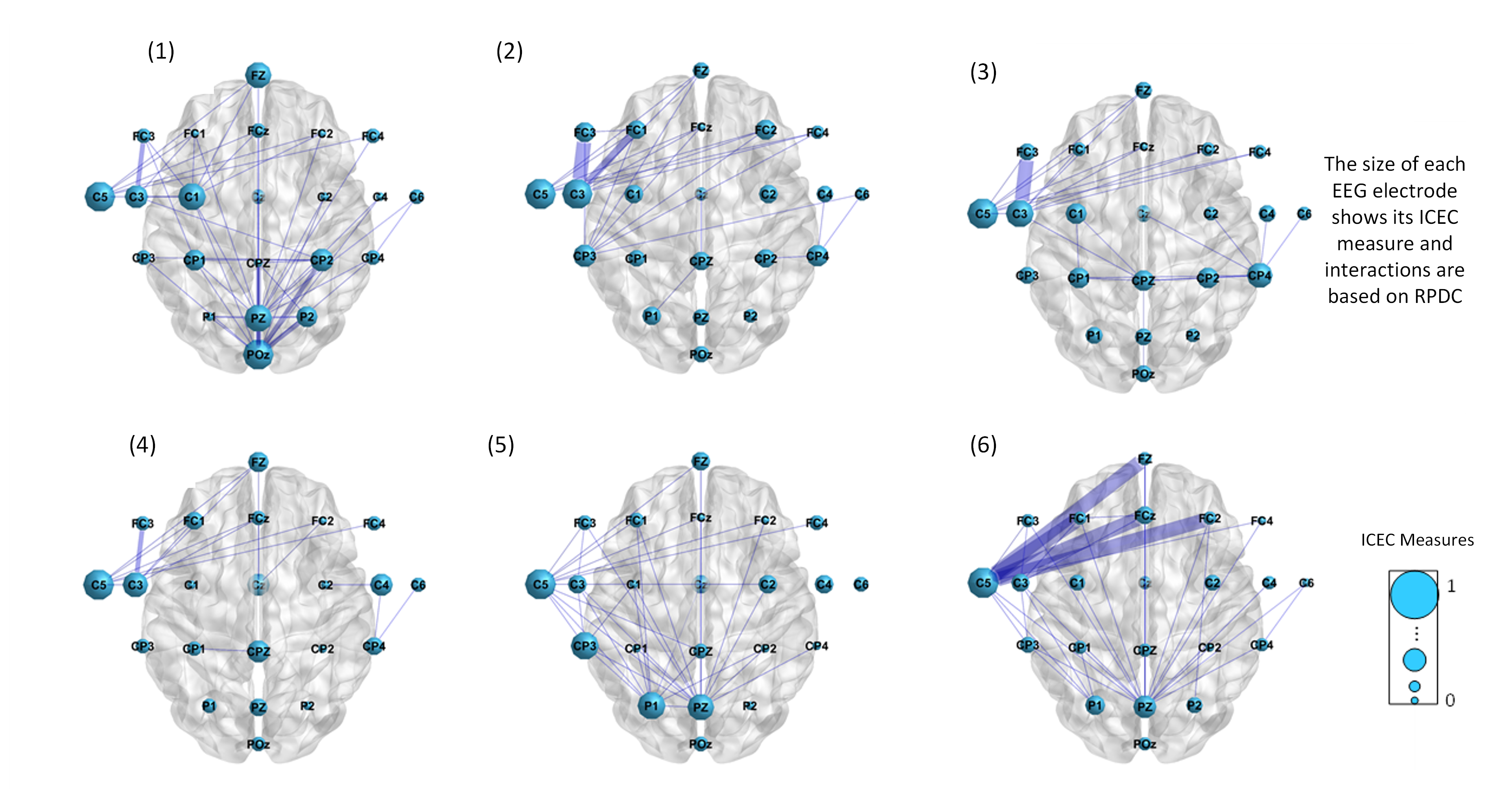}}
\caption{Effective Connectivity (RPDC) interactions among electrode nodes in the theta frequency range. The images depict the differences in RPDC among sensors every 500 ms within a 3-second timeframe of a single trial (user a03, Dataset 1).
}
\label{conn}   
\end{figure}

Figure~\ref{conn} visualizes the Effective Connectivity (RPDC) interactions among EEG electrode nodes within the theta frequency range for a single trial (user a03, Dataset 1). Each subplot represents the connectivity patterns at 500 ms intervals over a 3-second timeframe. The size of each electrode indicates its ICEC measure, highlighting the most influential electrodes in terms of effective connectivity. The strength and direction of the connections are represented by the links between the nodes, with thicker and darker lines indicating stronger interactions. The visualizations illustrate the temporal evolution of key electrode interactions during the trial.

\begin{figure}[ht]
    \centering
    \begin{subfigure}[t]{0.45\textwidth} 
        \caption*{A}
        \centering
        \includegraphics[width=\textwidth]{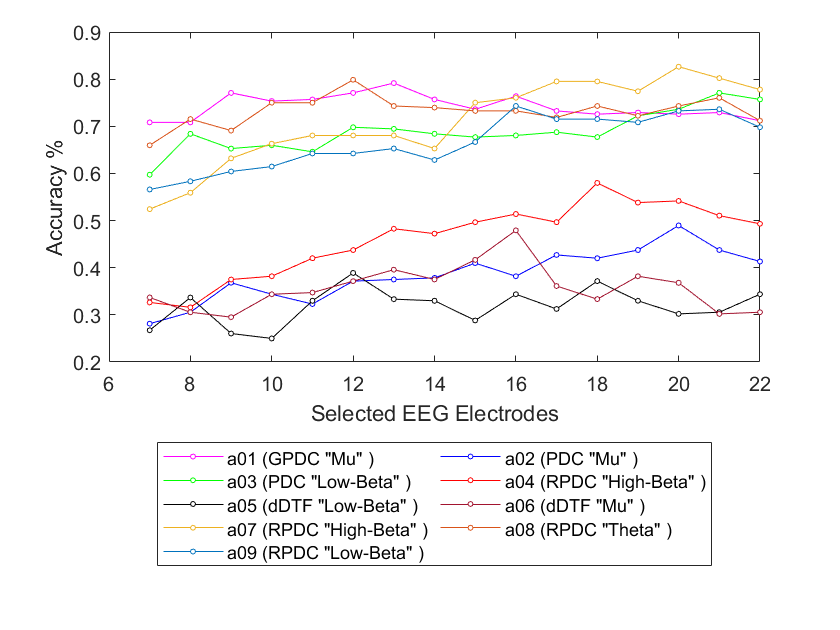} 
        \caption*{C}
        \label{fig:sub_a}
    \end{subfigure}
    \begin{subfigure}[t]{0.45\textwidth} 
        \caption*{B}
        \centering
        \includegraphics[width=\textwidth]{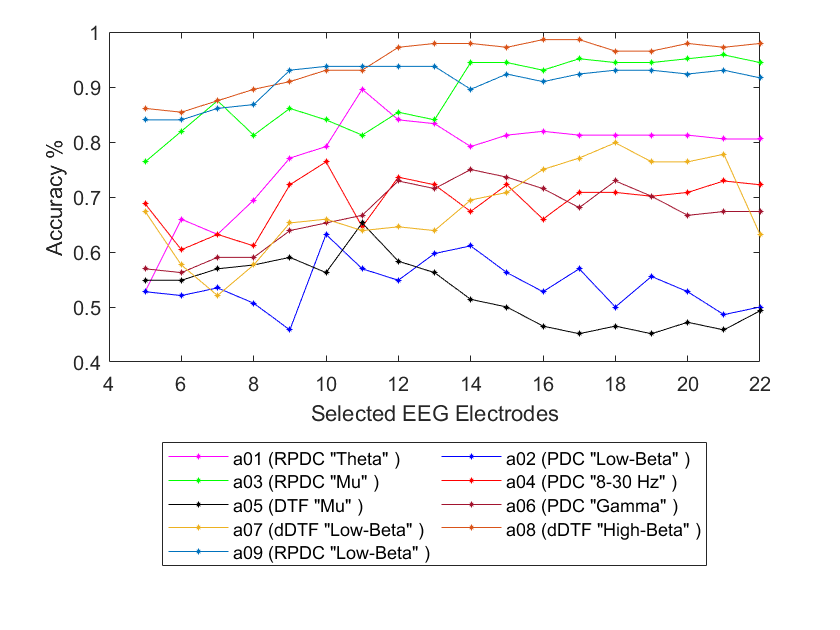} 
        \caption*{D} 
        \label{fig:sub_b}
    \end{subfigure}
    
    \begin{subfigure}[t]{0.45\textwidth} 
        \centering
        \includegraphics[width=\textwidth]{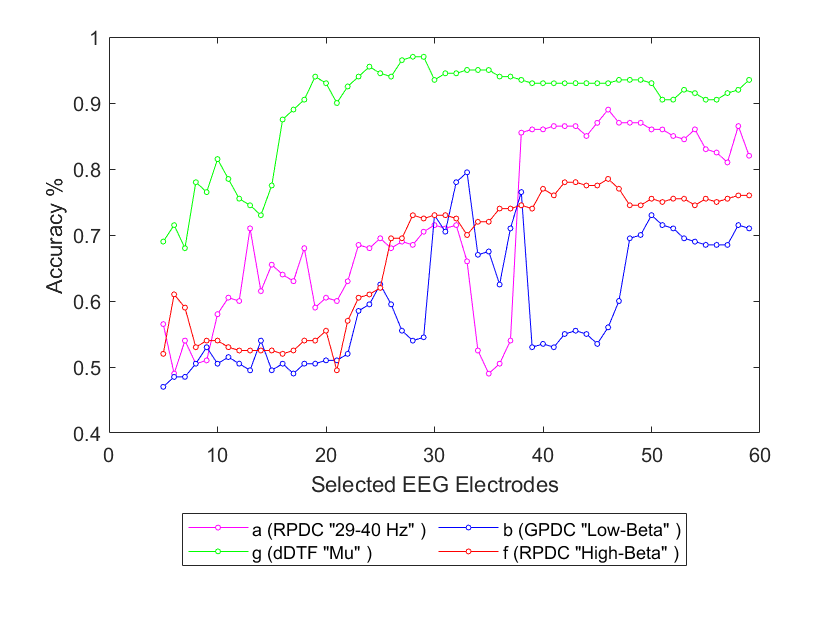} 
        \label{fig:sub_c}
    \end{subfigure}
    \begin{subfigure}[t]{0.45\textwidth} 
        \centering
        \includegraphics[width=\textwidth]{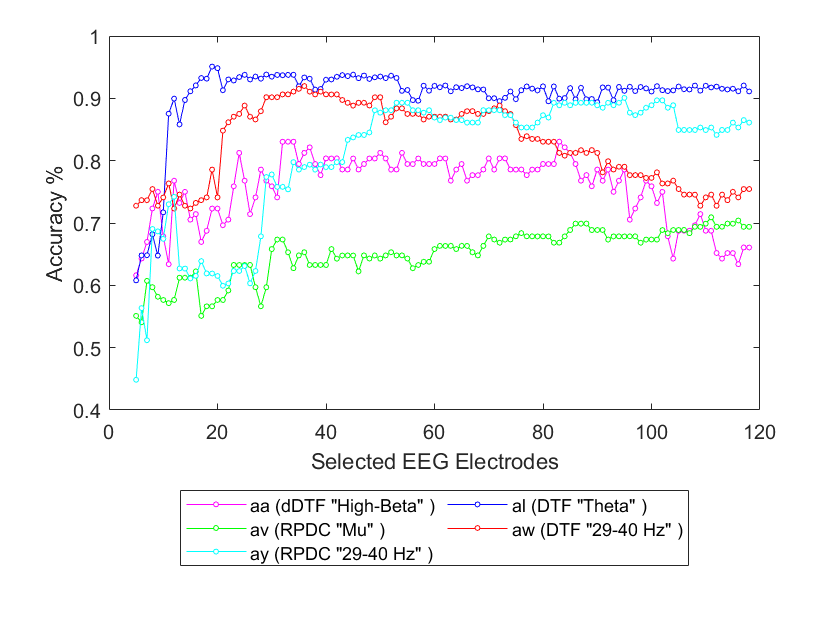} 
        \label{fig:sub_d}
    \end{subfigure}
    
    \caption{The accuracies associated with selected channels using our proposed method for dataset 1 (2-class (b) \& 4-class (a) tasks), dataset 2 (c), and dataset 3 (d). Details of frequency ranges and effective connectivity metrics used are provided in the legends. }
    \label{line chart}
\end{figure}

We examine the effect of increasing the number of selected electrodes based on the ICEC measures on classification accuracy across datasets and tasks (please see Figure~\ref{line chart}). The results show that after selecting a certain number of electrodes, accuracy significantly improves and then continues to increase slightly, maintaining consistent performance. This trend highlights the effectiveness of selecting key electrodes identified by the ICEC criterion in achieving optimal classification outcomes. For example, in Figure~\ref{line chart})(C), for participant ``g,'' the accuracy starts at approximately 0.6 with 10 selected electrodes and increases significantly to around 0.85 when 30 electrodes are selected. Beyond this point, the accuracy stabilizes with only slight improvements as more electrodes are added. Similarly, in Figure~\ref{line chart})(D), for participant ``aa,'' the accuracy begins at around 0.5 with 10 selected electrodes, rises significantly to approximately 0.9 by the time 40 electrodes are selected, and remains stable with minor fluctuations thereafter. These trends highlight that selecting key electrodes not only improves classification performance but also increases accuracy while significantly reducing the number of channels, thereby lowering data dimensionality.

\subsection{Comparison with other methods}

To evaluate our proposed method, we compared its results on three well-known global datasets with other state-of-the-art CSP-based methods. These results are shown in Tables 1 and 2. In Table 1, we compared the proposed method applied to dataset 1 (right hand vs left hand) with 3 other channel selection methods (SCSP1, SCSP2, IBGSA) and with the results of 2 other classification methods that don't make use of channel selection. The proposed method demonstrates competitive performance compared to other techniques across various subjects. While the proposed method achieves the highest mean accuracy of $82\% \pm 13.1\%$ with a lower standard deviation in selected channels (13), it closely aligns with methods like SCSP1 ($81.63\% \pm 13.7\%$), SCSP2 ($79.1\% \pm 15.6\%$), and surpasses IBGSA ($76.25\% \pm 10\%$), as well as \cite{34} ($74.5\% \pm 15.1\%$) and \cite{33} ($79.93\% \pm 14.1\%$). Notably, the reduced channel count highlights the efficiency of the proposed approach in maintaining high accuracy while simplifying the model, underscoring the robustness of the proposed method in optimizing classification accuracy with fewer resources.


\begin{sidewaystable}[htbp]
	\caption{Comparison of results of the proposed method applied to dataset 1 (Right hand vs Left hand) with 5 other methods. The numbers in parentheses indicate the number of selected channels, and * denotes a statistically significant difference (p-value < 0.05) based on a paired t-test. Standard deviations are also presented alongside the average accuracies.}
    \LARGE
	\begin{center}
		\resizebox{\textwidth}{!}{%
			\begin{tabular}{ccccccc} 
				\toprule
				\textbf{\textit{Subjects}} & 
				\textbf{\textit{Proposed}} & 
				\textbf{\textit{SCSP1}} & 
				\textbf{\textit{SCSP2}} & 
				\textbf{\textit{IBGSA}} & 
				\textbf{\textit{P. Gaur}} & 
                \textbf{\textit{Reza. H}} \\
				\midrule
                \textbf{\textit{}} & 
				\textbf{\textit{Method}} & 
				\textbf{\textit{\citet{31}}}& 
				\textbf{\textit{\citet{31}}}& 
				\textbf{\textit{\citet{32}}}& 
				\textbf{\textit{\citet{33}}}& 
				\textbf{\textit{\citet{34}}}\\ 
				\midrule
                \hline
				a01 & 89.58 (11) & \textbf{91.66} (13)& 91.66 (13) & 76.66 (9) & 91.49 & 90.28 \\ 
				a02 & 63.19 (11) & 67.36 (9) & 60.41 (4) & \textbf{76.66 }(6)& 60.56& 54.17 \\ 
				a03 & 95.83 (16) & \textbf{97.91} (14)& 97.14 (12) & 73.33 (7) & 94.16& 95.14 \\ 
				a04 & 76.39 (10)& 72.22 (14) & 70.83 (11) & 73.33 (8) & \textbf{76.72}& 65.97 \\ 
				a05 & 65.28 (11& 65.27 (11) & 63.19 (9) & \textbf{80} (11)& 58.52 & 61.11 \\ 
				a06 & \textbf{75} (14)& 66.67 (14) & 61.11 (10) & 73.33 (12) & 68.52& 65.28 \\ 
				a07 & 80 (19) & \textbf{84.72} (19)& 78.47 (15) & 76.66 (15) & 78.57 & 61.11  \\ 
				a08 & \textbf{98.91 }(16)& 97.22 (15) & 95.13 (15) & 80 (11) & 97.01 & 91.67 \\ 
				a09 & \textbf{93.95 }(10)& 91.66 (10) & 93.75 (5) & - & 93.85 & 86.11 \\ 
				\midrule
				\textbf{Mean} & \textbf{82 ± 13.1 (13)} & 81.63 ± 13.7 (13.2) & 79.1 ± 15.6 (9.3) & 76.25 ± 10* & 79.93 ± 14.1 (22) & 74.5 ± 15.1 (22)* \\  
				\bottomrule
                \hline
			\end{tabular}
		}
		\label{tabel_data1}
	\end{center}
\end{sidewaystable}

Table 2 gives the results of using our proposed EEG channel selection method on datasets 2 and 3 and a comparison with 4 other methods (SFFS, Improved SFFS, CSP-Rank, and BCS-CSP). As can be seen from the table, the proposed method (result 2) achieves the highest mean accuracy across both datasets (86.01\% $\pm$ 7.5\% for Dataset 2 and 87.56\% $\pm$ 7.90\% for Dataset 3) compared to other methods. For the proposed method (result 2), most subjects demonstrated higher accuracies. For instance, in Dataset 2, subjects ``a'' and ``g'' achieved accuracies of 89\% (46 channels) and 97\% (29 channels), respectively, which are significantly higher than other methods. Similarly, in Dataset 3, subjects ``aa'' and ``av'' achieved accuracies of 84.03\% (33 channels) and 96.43\% (105 channels), respectively. The proposed method (result 2) significantly outperforms Improved SFFS (67.25\% $\pm$ 3.49\%, $p\text{-value} < 0.05$) and SFFS (66.75\% $\pm$ 9.83\%, $p\text{-value} < 0.05$) in Dataset 2, while also surpassing CSP-Rank (79.1\% $\pm$ 6.21\%) and BCS-CSP (83.75\% $\pm$ 6.03\%). Similarly, in Dataset 3, the proposed method (result 2) shows statistically significant improvements ($p\text{-value} < 0.05$) over Improved SFFS (83.32\% $\pm$ 11.02\%) and SFFS (84.22\% $\pm$ 9.51\%) and maintains better performance than CSP-Rank (86.3\% $\pm$ 6.26\%) and BCS-CSP (86.34\% $\pm$ 8.30\%). These results emphasize the robustness of the proposed method in achieving superior classification performance, with most subjects showing significant improvements ($p\text{-value} < 0.05$) and a reduced number of selected channels.
\begin{sidewaystable}[htbp]
	\caption{Comparison of the proposed method's results with varying numbers of selected channels and their relative accuracies, applied to datasets 2 and 3 alongside five other methods. The numbers in parentheses indicate the number of selected channels, and * denotes a statistically significant difference (p-value < 0.05) based on a paired t-test. Standard deviations are also presented alongside the average accuracies.}
    \Huge
	\begin{center}
            
		\resizebox{\textwidth}{!}{%
			\begin{tabular}{ccccccccc}
				\toprule
				Datasets &\textbf{\textit{Subjects}} &
				\textbf{\textit{Proposed Method}} & 
				\textbf{\textit{Proposed Method}}&
				\textbf{\textit{Improved SFFS}} & 
				\textbf{\textit{SFFS}} & 
				\textbf{\textit{CSP-Rank}} & 
				\textbf{\textit{BCS-CSP}}  &
                \textbf{\textit{Regularized-CSP}}
                \\
				\midrule                    
				&\textbf{\textit{}} &  
				\textbf{\textit{result 1 }} & 
				\textbf{\textit{ result 2}}&
			  \textbf{ \textit{(2016)\citet{30}}} & 
				\textbf{\textit{(2016)\citet{30}}} & 
				\textbf{\textit{(2020)\citet{29}}} & 
				\textbf{\textit{(2020)\citet{29}}}&
                \textbf{\textit{(2024)\citet{2024}}}\\ 
				\midrule
                    \hline
				&a & 83.55 (21) & \textbf{89} (46) & 69 (6) & 60 (9) & 73 (5) & 78.5 (12)& 78 (16) \\
				&b & 79.53 (33) & \textbf{79.53} (33) & 63 (15) & 66 (19) & 77 (38) & 77.5 (40)& 72 (11)  \\ 
				Dataset 2&f & 76.51 (39) & 78.5 (46) & 65 (8) & 58 (19) & 89.5 (47) & 92 (33)& 65 (18)  \\ 
				&g & 96.1 (23) & \textbf{97} (29) & 72 (22) & 83 (21) & 77 (6) & 87 (13)& 75 (15)  \\ 
				\midrule
				&\textbf{Mean} & 83.93(25)±7.4 & \textbf{86.01}(29)±7.5 & 67.25 (12.7)± 3.49* & 66.75 (17.0) ± 9.83*& 79.1(24)± 6.21* & 83.75 (24.5) ± 6.03 & 72.5 (15)± 5.57* \\ 
				\midrule
				&aa & 82.1 (19) & \textbf{84.04} (33) & 76.4 (27) & 78.3 (26) & 81.1 (80) & 82.1 (105)&-  \\ 
				&al & 94.64 (14) & \textbf{96.43} (16)& 94.3 (47) & 93.6 (43) & 94.3 (74) & 95 (35)&-  \\ 
				Dataset3&av & 70.41 (22) & \textbf{74} (105)& 65 (18) & 68.6 (18) & 71.1 (64) & 72.1 (72)&-  \\ 
				&aw & 91.07 (24) & \textbf{93} (36)& 89.5 (27) & 87.9 (20) & 92.5 (27) & 90.7 (68) &- \\ 
				&ay & 89.29 (53) & 90.29 (53)& 91.4 (35) & \textbf{92.7} (35) & 92.5 (68) & 91.8 (111)&-  \\ 
				\midrule
				&\textbf{Mean} & 85.50 (26.4) ± 8.58 & \textbf{87.56}(48.6) ± 7.90& 83.32(30.8)±11.02* & 84.22 (28.4) ± 9.51* & 86.3 (62.6) ± 8.93 & 86.34 (78.2) ± 8.30&-  \\ 
				\bottomrule
                    \hline  
			\end{tabular}
		}
		\label{table_data2and3}
	\end{center}
\end{sidewaystable}

\begin{figure}[htbp]
\centerline{\includegraphics[width=0.90\columnwidth]{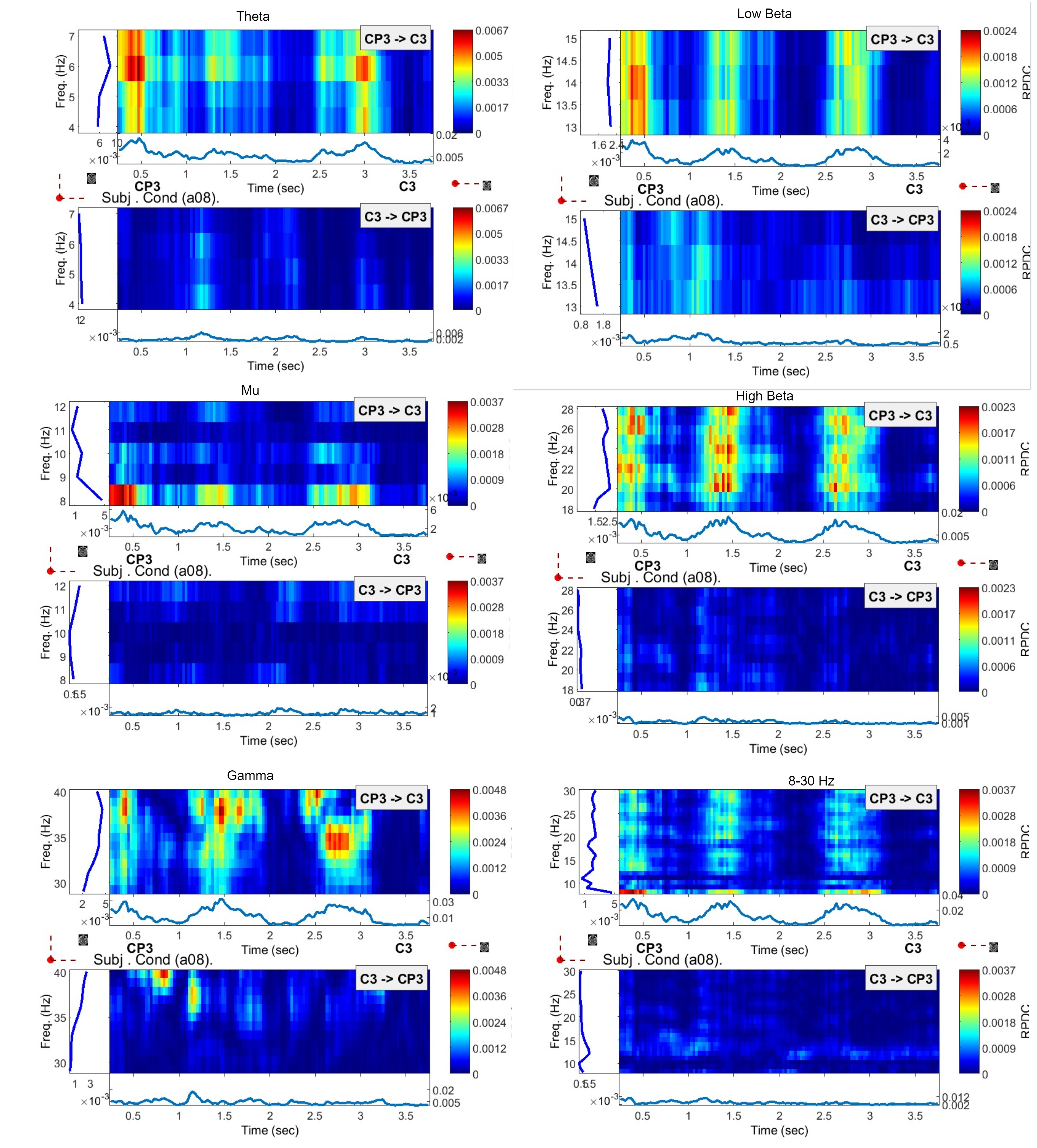}}
\caption{RPDC time-frequency illustrations of C3 and CP3 (which possess the highest ICEC amounts compared to other electrodes) for participant ``a08'', Dataset 1, across six frequency ranges.
}
\label{frequencies}   
\end{figure}

It is worth mentioning that, in our investigation, for a specific participant, electrodes with the highest ICEC amounts consistently maintained their ranks across all frequency ranges. For example, Figure~\ref{frequencies} illustrates RPDC time-frequency representations for the electrodes C3 and CP3, which exhibit the highest ICEC values among all electrodes for participant ``a08'' from Dataset 1. The plots depict interactions across six distinct frequency ranges, including Theta, Low Beta, Mu, High Beta, Gamma, and the broader 8--30 Hz band. These visualizations emphasize the strong and consistent effective connectivity between the selected electrodes, showcasing their dominant role in capturing relevant neural dynamics within the analyzed frequency bands.

\subsection{Potential Applications of ICEC Criterion}

In this study, we employed the ICEC criterion to quantify the intensity of effective connectivity (EC) in each node (channel). We demonstrated how this criterion can be applied for channel selection, enabling improved system performance while reducing the number of required channels. Additionally, we propose that the ICEC criterion has potential applications in other areas. For instance, in previous studies investigating the impacts of language-based circumstances on infants’ brains and also fatigue pattern recognition \cite{45, abdollahpour6effective}, we utilized the ICEC criterion and another type of graph construction method for data visualization. In the present study, we further defined Channel Edge and Weight matrices based on this criterion, specifically applying it to quantify EC in functional Near Infrared Spectroscopy (fNIRS) data. We posit that the ICEC criterion can be broadly applied in contexts where either effective or functional connectivity is of interest, particularly in scenarios requiring decisions based on the degree of interaction between elements.

\section{Conclusion}
In this study, we introduced the "Importance of Channels based on Effective Connectivity" (ICEC) criterion, a novel approach for quantifying effective connectivity (EC) in individual channels. Leveraging this criterion, we proposed an unsupervised channel selection method that considers the intensity of interactions among channels, leading to optimized sensor selection for classification tasks. Our method was rigorously evaluated on three widely recognized datasets across four categories, employing five effective connectivity metrics (PDC, GPDC, RPDC, DTF, and dDTF) and seven frequency ranges for comprehensive analysis. The results revealed consistent performance enhancements across all categories, with significant reductions in the number of selected electrodes, thereby demonstrating the efficiency of our approach. Furthermore, our method outperformed traditional techniques such as SFFS, CSP-Rank, and BCS-CSP, achieving superior classification accuracies of 82\%, 86.01\%, and 87.56\%, respectively. These outcomes validate the robustness and reliability of the ICEC-based channel selection strategy. Beyond improving classification outcomes, the ICEC criterion holds promise for broader applications, including data visualization and adaptability to other neuroimaging modalities. By optimizing sensor selection and enhancing feature extraction, this approach contributes to advancing neuroimaging analysis and related fields. Future work will explore its potential in diverse settings, further expanding its applicability and impact.
\section*{ACKNOWLEDGEMENTS}
This work was supported by Isfahan University of Medical Sciences under Grants 2401163.

 \bibliographystyle{elsarticle-num-names} 
 \bibliography{references}







\end{document}